\documentclass[twoside,twocolumn,9pt]{article}
\usepackage{extsizes}
\usepackage[super,sort&compress,comma]{natbib} 
\usepackage[version=3]{mhchem}
\usepackage[left=1.5cm, right=1.5cm, top=1.785cm, bottom=2.0cm]{geometry}
\usepackage{balance}
\usepackage{times,mathptmx}
\usepackage{sectsty}
\usepackage{graphicx} 
\usepackage{subcaption}
\usepackage{amssymb}
\usepackage{lastpage}
\usepackage[format=plain,justification=justified,singlelinecheck=false,font={stretch=1.125,small,sf},labelfont=bf,labelsep=space]{caption}
\usepackage{float}
\usepackage{fancyhdr}
\usepackage{fnpos}
\usepackage[english]{babel}
\usepackage{array}
\usepackage{droidsans}
\usepackage{charter}
\usepackage[T1]{fontenc}
\usepackage[usenames,dvipsnames]{xcolor}
\usepackage{setspace}
\usepackage[compact]{titlesec}

\usepackage[normalem]{ulem}

\newif\ifJCCHighlitedChanges
\def\ifJCCHighlitedChanges{\iftrue}
\ifJCCHighlitedChanges
  
  \def\SJCC#1{{\color{blue}\sout{#1}}}

\else
  
  \def\SJCC#1{\relax}
  
  \def\CJCP#1{\relax}
\fi

\newif\ifNKDHighlitedChanges
\def\ifNKDHighlitedChanges{\iffalse}
\ifNKDHighlitedChanges
  \def\INKD#1{{\color{purple}#1}}
  \def\SNKD#1{{\color{purple}\sout{#1}}}
  
\else
  \def\INKD#1{#1}
  \def\SNKD#1{\relax}
\fi

\definecolor{cream}{RGB}{222,217,201}

\begin{document}

\pagestyle{fancy}
\thispagestyle{plain}
\fancypagestyle{plain}{

\fancyhead[C]{\includegraphics[width=18.5cm]{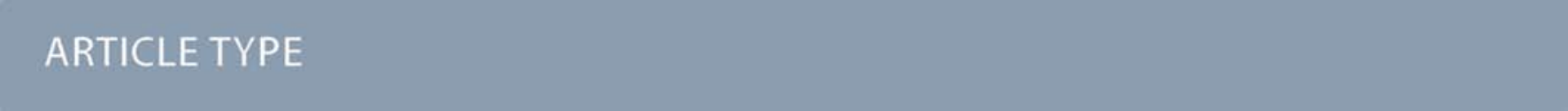}}
\fancyhead[L]{\hspace{0cm}\vspace{1.5cm}\includegraphics[height=30pt]{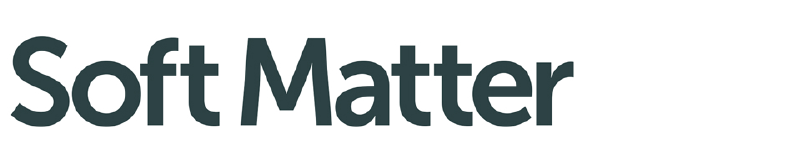}}
\fancyhead[R]{\hspace{0cm}\vspace{1.7cm}\includegraphics[height=55pt]{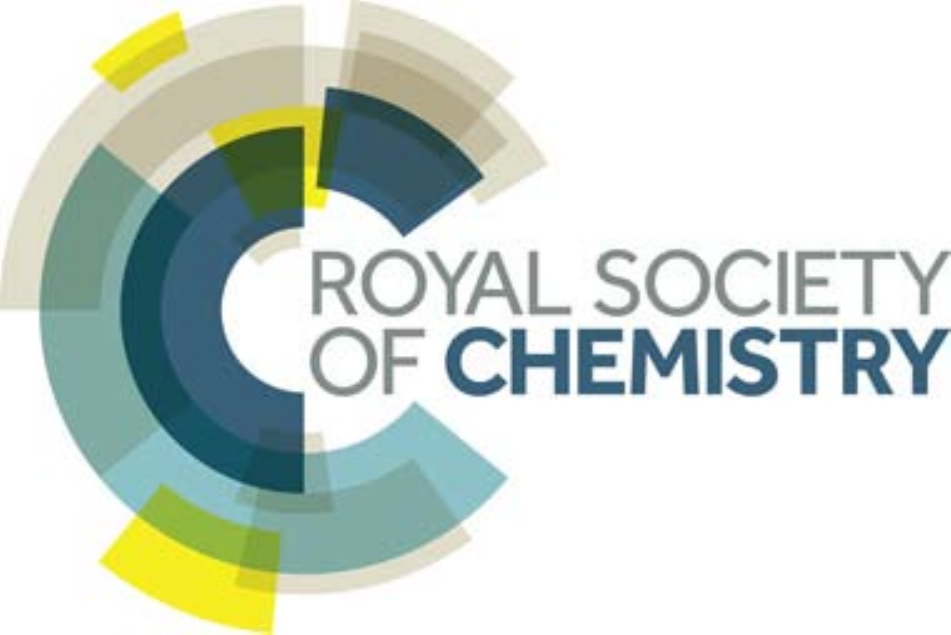}}
\renewcommand{\headrulewidth}{0pt}
}

\makeFNbottom
\makeatletter
\renewcommand\LARGE{\@setfontsize\LARGE{15pt}{17}}
\renewcommand\Large{\@setfontsize\Large{12pt}{14}}
\renewcommand\large{\@setfontsize\large{10pt}{12}}
\renewcommand\footnotesize{\@setfontsize\footnotesize{7pt}{10}}
\makeatother

\renewcommand{\thefootnote}{\fnsymbol{footnote}}
\renewcommand\footnoterule{\vspace*{1pt}%
\color{cream}\hrule width 3.5in height 0.4pt \color{black}\vspace*{5pt}} 
\setcounter{secnumdepth}{5}

\makeatletter 
\renewcommand\@biblabel[1]{#1}            
\renewcommand\@makefntext[1]%
{\noindent\makebox[0pt][r]{\@thefnmark\,}#1}
\makeatother 
\renewcommand{\figurename}{\small{Fig.}~}
\sectionfont{\sffamily\Large}
\subsectionfont{\normalsize}
\subsubsectionfont{\bf}
\setstretch{1.125} 
\setlength{\skip\footins}{0.8cm}
\setlength{\footnotesep}{0.25cm}
\setlength{\jot}{10pt}
\titlespacing*{\section}{0pt}{4pt}{4pt}
\titlespacing*{\subsection}{0pt}{15pt}{1pt}

\fancyfoot{}
\fancyfoot[LO,RE]{\vspace{-7.1pt}\includegraphics[height=9pt]{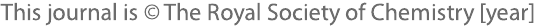}}
\fancyfoot[CO]{\vspace{-7.1pt}\hspace{13.2cm}\includegraphics{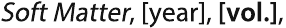}}
\fancyfoot[CE]{\vspace{-7.2pt}\hspace{-14.2cm}\includegraphics{head_foot/RF}}
\fancyfoot[RO]{\footnotesize{\sffamily{1--\pageref{LastPage} ~\textbar  \hspace{2pt}\thepage}}}
\fancyfoot[LE]{\footnotesize{\sffamily{\thepage~\textbar\hspace{3.45cm} 1--\pageref{LastPage}}}}
\fancyhead{}
\renewcommand{\headrulewidth}{0pt} 
\renewcommand{\footrulewidth}{0pt}
\setlength{\arrayrulewidth}{1pt}
\setlength{\columnsep}{6.5mm}
\setlength\bibsep{1pt}

\makeatletter 
\newlength{\figrulesep} 
\setlength{\figrulesep}{0.5\textfloatsep} 

\newcommand{\topfigrule}{\vspace*{-1pt}%
\noindent{\color{cream}\rule[-\figrulesep]{\columnwidth}{1.5pt}} }

\newcommand{\botfigrule}{\vspace*{-2pt}%
\noindent{\color{cream}\rule[\figrulesep]{\columnwidth}{1.5pt}} }

\newcommand{\dblfigrule}{\vspace*{-1pt}%
\noindent{\color{cream}\rule[-\figrulesep]{\textwidth}{1.5pt}} }

\makeatother

\twocolumn[
  \begin{@twocolumnfalse}
\vspace{3cm}
\sffamily
\begin{tabular}{m{4.5cm} p{13.5cm} }

\includegraphics{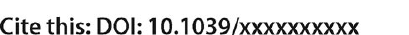} & \noindent\LARGE{\textbf{Bacterial Motility Enhances Adhesion to Oil Droplets$^\dag$}} \\
\vspace{0.3cm} & \vspace{0.3cm} \\

 & \noindent\large{Narendra K. Dewangan and Jacinta C. Conrad\textit{$^{\ast}$}} \\

\includegraphics{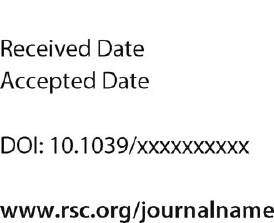} & \noindent\normalsize{Adhesion of bacteria to liquid-liquid interfaces can play a role in the biodegradation of dispersed hydrocarbons and in biochemical and bioprocess engineering. Whereas thermodynamic factors underpinning adhesion are well studied, the role of bacterial activity on adhesion is less explored. Here, we show that bacterial motility enhances adhesion to surfactant-decorated oil droplets dispersed in artificial sea water. Motile \textit{Halomonas titanicae} adhered to hexadecane droplets stabilized with dioctyl sodium sulfosuccinate (DOSS) more rapidly and at greater surface densities compared to nonmotile \textit{H.\ titanicae}, whose flagellar motion was arrested through addition of a proton \INKD{uncoupler}. Increasing the concentration of DOSS reduced the surface density of both motile and nonmotile bacteria as a result of the reduced interfacial tension.}

\end{tabular}

 \end{@twocolumnfalse} \vspace{0.6cm}

]

\renewcommand*\rmdefault{bch}\normalfont\upshape
\rmfamily
\section*{}
\vspace{-1cm}


\footnotetext{\textit{Department of Chemical and Biomolecular Engineering, University of Houston, Houston, TX 77204}}
\footnotetext{\textit{$^{\ast}$~E-mail: jcconrad@uh.edu}}
\footnotetext{\dag~Electronic Supplementary Information (ESI) available. See DOI: 10.1039.xxxxxx}

\section{Introduction}

\par Adhesion of bacteria at the interface between two liquids can alter the rate of biodegradation of hydrocarbons\cite{kostka2011hydrocarbon, abbasnezhad2011adhesion, mahmoudi2013rapid, bagby2017persistence} during marine oil spills and the efficacy of bioprocess engineering operations that involve multiple immiscible fluid phases.\cite{morrish2008enhanced, oda2020microbial} Adhesion of bacteria on solid or liquid surfaces can depend on physicochemical properties of the surfaces (e.g., surface functionality and topography) and liquids (e.g., surfactant concentration, ionic strength, pH, and concentration of the carbon source) as well as cell surface hydrophobicity (which, in turn, depends on adhesin and polysaccharide expression).\cite{marshall1971mechanism,bolster2001effect, de2007adhesion, chakraborty2010surface, dewangan2018adhesion} These properties affect the electrostatic and van der Waals forces that control the thermodynamics of adhesion for micron-size bacteria at interfaces. Many bacteria, however, are active and motile, moving randomly and in response to chemical\cite{berg1972chemotaxis, blair1995bacteria, berg2003rotary, harshey2003bacterial, szurmant2004diversity, conrad2018review} or physical (e.g., gravity,\cite{desai2019hydrodynamics} viscosity,\cite{shaik2018locomotion} flow\cite{wheeler2019not}) gradients. Thus, motility may also affect adhesion to liquid-liquid interfaces.

\par Bacteria motility is known to enhance adhesion of bacteria to solid surfaces.\cite{korber1994effect, kogure1998attachment, camesano1998influence, morisaki1999effect, vigeant2002reversible, de2007impact, kerchove2008bacterial, tamar2016role, jindai2020adhesion} For example, the fractional surface coverage by motile \textit{Pseudomonas\ aeruginosa} PAO1 bacteria is up to 2.5 times greater than that of a nonmotile mutant, depending on the ionic strength and Reynolds number.\cite{kerchove2008bacterial} This result suggests that motility enabled bacteria to attach to surface sites that were otherwise inaccessible; in the picture of Ref.\ \citenum{kerchove2008bacterial}, swimming provided a kinetic ``force'' that competed with the electrostatic and hydrodynamic forces controlling attachment of nonmotile cells.\cite{kerchove2008bacterial} Similarly, approximately five to ten times more motile \textit{Escherichia coli} bacteria adhered on a nanostructured biocidal surface compared to mutants that lacked rotating flagella or the receptors required for chemotaxis.\cite{jindai2020adhesion} Motility may increase the frequency at which bacteria collide with the surface, as suggested by the positive correlation between attachment rate and swimming speed for motile \textit{Alcaligenes} and \textit{Alteromonas spp}.\cite{morisaki1999effect} As a final example, motile and nonmotile \textit{E.\ coli} bacteria were located on average at different distances above the surface, affecting the interactions which acted upon the bacteria and hence their surface attachment.\cite{vigeant2002reversible} Motility is less studied near liquid-liquid interfaces, such as the oil-water interfaces encountered in an oil spill scenario. Recent studies reveal that \textit{P.\ aeruginosa} bacteria display a variety of motility behaviors near a planar oil-water interface,\cite{vaccari2015films, vaccari2018cargo, deng2020motile} and that bacterial motility can provide sufficient energy to move dispersed oil droplets.\cite{dewangan2019rotating,ramos2020bacteria} How bacterial motility affects adhesion to oil-water interfaces, however, remains incompletely understood.

\par Here, we show that bacteria motility enhances adhesion of the marine bacterium \textit{Halomonas titanicae} on hexadecane droplets suspended in artificial seawater. Using confocal microscopy and single cell tracking algorithms, we quantified the number of cells adhering to hexadecane droplets over time. \INKD{To render cells nonmotile, we} added  a proton \INKD{uncoupler}, \INKD{carbonyl cyanide m-chlorophenyl hydrazone} (CCCP), \INKD{or applied mechanical shear to remove the flagella}. Both motile and nonmotile cells exhibited first-order Langmuir kinetics for adhesion. The time constant extracted from the Langmuir fit was smaller for motile bacteria. Furthermore, the long-time density of bacteria on the droplet was greater for motile bacteria. Increasing the concentration of an anionic surfactant, diocytl sodium sulfosuccinate (DOSS), reduced the oil-water interfacial tension, causing fewer cells to attach to the interface. The long-time density of motile bacteria on the oil-water interface was greater than that of the nonmotile bacteria over a wide range of DOSS concentrations. Our results suggest that motility can lead to faster and greater adhesion of bacteria, which may advantage bacteria during biodegradation and other processes that involve access to a dispersed fluid phase.

\section{Materials and Methods}
\subsection{Chemicals} 
\par Zobell marine broth 2216 (HiMedia lab), sodium pyruvate (Amresco), nutrient agar (Difco), hexadecane ($\geq$ 99\%, Sigma-Aldrich), dioctyl sodium sulfosuccinate (DOSS, $\geq$ 97\%, Sigma-Aldrich), SYTO9 (ThermoFisher), carbonyl cyanide 3-chlorophenylhydrazone (CCCP, $\geq$ 97\%, Sigma-Aldrich), sodium chloride ($\geq$ 99\%, BDH), magnesium chloride hexahydrate ($\geq$ 99\%, Alfa Aesar), magnesium sulfate heptahydrate ($\geq$ 99.5\%, Sigma-Aldrich), calcium chloride dihydrate (Sigma-Aldrich), potassium chloride ($\geq$ 99\%, BDH), potassium nitrate ($\geq$ 99\%, EMD), dipotassium phosphate ($\geq$ 98\%, Sigma-Aldrich), ethylene glycol ($\geq$ 99.8\%, Sigma-Aldrich), and diiodomethane ($\geq$ 99, Sigma-Aldrich) were used as received.

\subsection{Bacteria Strains and Growth Conditions} 
\par We studied two species of marine bacteria. The Bead 10BA strain is closely related to \textit{Halomonas titanicae}. It was isolated from samples collected at 1509 m during the Deepwater Horizon oil spill in the Gulf of Mexico by Dr.\ Romy Chakraborty and Dr.\ Gary Anderson (Lawrence Berkeley National Laboratory) and recieved from Dr.\ Douglas Bartlett (Scripps Institute of Oceanography, UCSD). Hereafter, it is referred to by its closest species name, \textit{H.\ titanicae}. \textit{H.\ titanicae} is moderately halophilic, Gram-negative, and rod-shaped, with diameter of 0.5 $\--$ 0.8 $\mu$m and length of 1.5 $\--$ 6 $\mu$m. These bacteria swim using their peritrichous flagella.\cite{lee2015halomonas} \textit{Marinobacter hydrocarbonoclasticus} (ATCC 49840) was obtained from Dr.\ Bartlett. \textit{M. hydrocarbonoclasticus} is halotolerant, Gram-negative, and rod-shaped, with length of 2 $\--$ 3 $\mu$m and diameter of 0.3 $\--$ 0.6 $\mu$m.\cite{gauthier1992marinobacter} It is nonmotile under the conditions of our experiments. Marine agar plates (38.7 g L$^{-1}$ marine broth, 10 g L$^{-1}$ sodium pyruvate, 15 g L$^{-1}$ agar) were streaked from frozen stocks of these bacteria and incubated at 30 $^\circ$C for 40 h. \INKD{To initiate the principal culture,} 20 mL of culture medium (38.7 g L$^{-1}$ marine broth and 10 g L$^{-1}$ pyruvate) was inoculated from a single colony of bacteria and incubated for 20 h in an orbital incubator shaker (New Brunswick Scientific) at 200 rpm and 30 $^\circ$C. To prepare subcultures, 20 mL of culture medium was inoculated with 65 $\mu$L of the principal culture and grown to late exponential phase in an orbital incubator shaker at 30 $^\circ$C and 200 rpm for 20 h.

\subsection{Zeta Potential} 
\par For zeta potential measurements, bacteria cells were grown to late exponential phase. \INKD{First,} 20 mL of each bacteria suspension was centrifuged at 5000 \textit{g} for 10 minutes in a Sorvall ST 16 Centrifuge (ThermoFisher Scientific). The pellet was resuspended in 20 mL MilliQ water and centrifuged again. After repeating this process one more time, the pellet was resuspended in MilliQ water. The resultant suspension was diluted to an optical density at 600 nm (OD$_\mathrm{600 nm}$) of 0.04 $\--$ 0.06 (Laxco DSM-Micro Cell Density Meter) with MilliQ water. The zeta potentials of suspensions were measured using a Nicomp 380 $\zeta$-potential analyzer (Table S1).\cite{sharma2016subnanometric}

\subsection{Contact Angle and Surface Energy} 
\par Bacteria suspensions for surface measurements were prepared in a similar manner as those for zeta potential measurements, except that the final OD was adjusted to 1.0. Each bacteria suspension was filtered through cellulose acetate membrane filters (pore diameter 0.45 $\mu$m, Advantec) under vacuum at 100 mm Hg below atmospheric pressure using a GEM 8890 vacuum pump (Welch) to create a bacterial lawn.\cite{mclay2017level, yadav2017tuning} Contact angles were measured for three liquids (MilliQ water, ethylene glycol and diiodomethane) on the lawns using a Dataphysics OCA 15EC goniometer. The surface energy of the bacteria was calculated using the method of Wu.\cite{wu1971calculation,wu1973polar}

\subsection{Interfacial Tension} 
\par The interfacial tension of hexadecane/water was measured as a function of DOSS concentration via the pendant drop method using a Dataphysics OCA 15EC goniometer. The outer phase was artificial seawater (ASW: 0.33 M NaCl, 0.06 M MgCl$_{2}$, 0.03 M MgSO$_{4}$, 0.016 M CaCl$_{2}$, 0.007 M KCl, 0.019 M KNO$_{3}$, and 0.0007 M HK$_{2}$PO$_{4}$) \cite{abbasi2018attachment, godfrin2018behavior}  with DOSS (1 $\--$ 500 ppm) and the inner phase was hexadecane.  
\subsection{Imaging Chamber for Confocal Microscopy} 
\par Two types of chambers were prepared for imaging experiments. For nonmotile cells, a glass slide and a glass cover slip were exposed to oxygen plasma for 2 min. A PDMS layer of thickness 1 $\--$ 1.5 mm on a 100 mm silicon wafer was prepared by spincoating at 100 rpm for 30 seconds using a spin coater (Brewer Science CEE 200CB). A 15 $\times$ 20 mm${^2}$ rectangle of PDMS was cut from the layer and a well of 3 $\--$ 4 $\times$ 6 $\--$ 8 mm${^2}$ was created to contain the bacteria sample. The PDMS rectangle was placed onto a plasma-cleaned glass slide right after the plasma treatment and covered with a plasma-cleaned glass cover slip (Fig.\ S1). A similar chamber was prepared for motile cells, except that the top cover slip was replaced with a cellulose dialysis tubing patch (12 -- 14 kDa cutoff, Carolina Biological) attached with vacuum grease. For motile cell experiments, NaCl solution was introduced through the membrane to reduce cell motility during imaging.

\subsection{Turning Off Motility} 
\par Many bacteria can swim using one or more flagella, which are driven by electrochemical gradients of protons or sodium ions across the cytoplasmic membranes.\cite{berg2003rotary} Carbonyl cyanide 3-chlorophenylhydrazone (CCCP) was introduced to cell suspensions to inhibit cell motility. CCCP is a protonphore that collapses the proton motive force (PMF) across the cytoplasmic membrane, and thereby halts cell motility.\cite{gosink2000requirements} A solution of CCCP of concentration 5 mM was added to bacterial suspensions at an appropriate volume to obtain a final concentration of 5 $\mu$M and incubated for about 3 minutes. Using a brightfield microscope, we confirmed that nearly all cells stopped swimming after 3 minutes of incubation time.

\par We also mechanically sheared off the flagella by rapidly agitating 50 mL of a washed cell suspension at OD 0.4 using a blender (Oster 6642) in liquify mode at high for 15 sec. This process temporarily renders cells nonmotile by removing flagella. \INKD{Sheared cells on average did not recover motility over 2 h (Fig. S1).} \cite{eisenbach1981bacterial} Sheared cells were stained and imaged using the protocol for nonmotile cells described in the following subsection.

\subsection{Imaging Cells Adhering on Hexadecane Droplets} \par From a suspension of cells grown to late exponential phase, 20 mL was centrifuged at 2000 \textit{g} for 10 minutes. After the supernatant was discarded, the resultant pellet was suspended in 2x ASW by gentle shaking and the OD was adjusted to 0.4. SYTO 9 (1 $\mu$L) was added to 1 mL of this cell suspension and incubated at room temperature in the dark for 2 $\--$ 5  minutes.

\par For nonmotile bacteria experiments, 1 $\mu$L of 5 mM CCCP was added along with 1 $\mu$L of SYTO9 to 1 mL of cell suspension, which resulted in a final CCCP concentration of 5 $\mu$M. Hexadecane-in-water emulsions were prepared by manually shaking 10 $\mu$L of hexadecane in 1 mL of MilliQ water containing DOSS (4, 40, 100, 200, 300, and 400 ppm) in a 1.5 mL Eppendorf tube. From the cell suspension, 200 $\mu$L was transferred into a 1.5 mL Eppendorf tube, to which 200 $\mu$L of emulsion was subsequently added. This protocol resulted in a final CCCP concentration of 2.5 $\mu$M. The resulting suspension (OD = 0.2, 5.4$\times10^8$ cells mL$^{-1}$) was introduced in an imaging chamber and sealed using vacuum grease.

\par \INKD{Because \textit{H.\ titanicae} bacteria  remained motile at the droplet surface after adhesion, we halted the motility by addition of NaCl to enable  quantification of cells.} \INKD{Although cells adhered on the droplet could move along the its surface, they did not appear to detach.} A modified protocol was adopted for imaging suspensions containing motile cells. For these experiments, 10 $\mu$L of NaCl solution (350 g L$^{-1}$ NaCl in MilliQ water) was introduced into a suspension of motile cells through the membrane that formed the top boundary of the imaging chamber. Sodium chloride diffused through the membrane into the cell-and-emulsion suspension and the cells stopped swimming \INKD{in 1 to} 3 minutes due to the high NaCl content. Suspensions were imaged within 15 min after introduction of NaCl. The cessation of swimming motion facilitated imaging and quantification because cells exhibited minimal displacement during the 3-D scanning.

\par Bacteria were imaged in 3-D over time using a VT-Infinity (Visitech, Sunderland, UK) confocal scanhead. The confocal scanhead was mounted on an inverted microscope (Leica Microsystems DM4000) bearing a 40X oil-immersion lens (HCX PL APO, NA 1.25 -- 0.75). Stacks of two-dimensional images separated by a height $\delta z = 0.31$ $\mu$m were acquired at times t = 5, 10, 20, 30, 60, 90, and 120 min after the cell suspension was added to the emulsion. \INKD{For imaging motile bacteria, 350 g L$^{-1}$ NaCl was introduced at 5, 10, 20, 30, 60, 90, and 120 min after the bacteria suspension was mixed with the emulsion. Subsequently, z-stacks were acquired after cell motility was arrested (between 1 and 3 min) after introduction of NaCl.} Each experiment was repeated with at least four independent cultures. For equilibrium adhesion experiments, 3-D images were acquired 1 hour after the emulsion was introduced into the cell suspension. Cells at the oil-water interface were enumerated using both particle tracking algorithms written in MATLAB\INKD{ \cite{dewangan2018adhesion}} and through manual counting using ImageJ. The cell density was calculated by dividing the number of cells adhered on the top hemisphere of a droplet by the surface area of the hemisphere.

\section{Results and Discussion}

\subsection{Bacterial Adhesion on Hexadecane Droplets Over Time.} 

\par
\begin{figure}[!ht]
\centering
\includegraphics[width=3.25in]{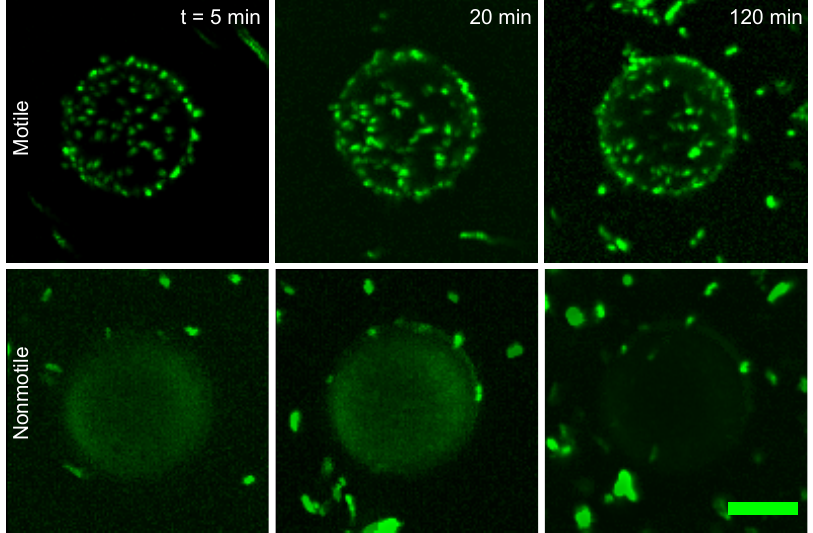}
\caption{\label{fig:images_vst} 2D projections of 3D confocal images of cells adhered on 20 $\mu$m hexadecane droplets for motile bacteria (top panel), and nonmotile bacteria (bottom panel) at $t =$ 5 (first column), 20 (second column), and 120 min (third column). DOSS concentration is 2 ppm. Scale bar is 10 $\mu$m. }
\end{figure}

\begin{figure}[!ht]
\centering
\includegraphics[width=3.25in]{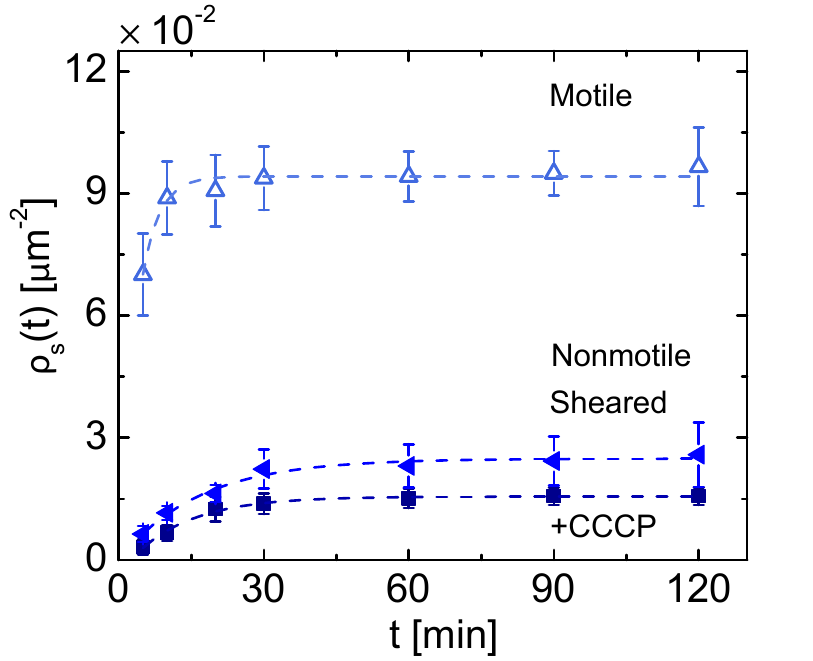}
\caption{\label{fig:rho_vst} Interfacial density $\rho_s(t)$ of cells adhered on 20 $\mu$m hexadecane droplets as a function of time for\SNKD{motile and nonmotile} \INKD{motile (-CCCP, open triangles), chemically nonmotile (+CCCP, solid squares), and mechanically nonmotile (sheared, solid left triangles)} \textit{H.\ titanicae}. The DOSS concentration in suspension was 2 ppm. Dashed lines represent first-order Langmuir adsorption fits. Error bars represent the standard deviation calculated from at least 30 droplets (at least 6 droplets for at least 5 independent bacterial cultures).} 
\end{figure}

 Using confocal microscopy and single-cell tracking, we analyzed the adhesion of \textit{Halomonas titanicae} bacteria over time on 20 $\mu$m hexadecane droplets stabilized by DOSS and suspended in ASW. We assessed the effects of motility on adhesion by comparing results from suspensions in the absence and presence of CCCP, which arrests cell motility by collapsing the PMF.\cite{gosink2000requirements} Examination of confocal micrographs reveals that the number of bacteria adhered on hexadecane droplets changes over time (Fig.\ \ref{fig:images_vst}).  The density of cells at the oil-water interface $\rho_\mathrm{s}$, calculated from the 3-D images, initially increases with time and is approximately constant after 60 minutes for both motile and nonmotile bacteria (Fig. \ref{fig:rho_vst}). \INKD{The interfacial cell density for motile bacteria is higher than that of both the chemically modified and mechanically sheared nonmotile bacteria. Although $\rho_s(t)$ is slightly greater for the mechanically sheared bacteria than the chemically modified bacteria on all time scales, the difference is within the measurement error.}  The time-dependent densities of both motile and nonmotile bacteria follow a Langmuir first-order kinetic model, $\rho_\mathrm{s}(t) = \rho_\mathrm{s,\infty} - (\rho_\mathrm{s,\infty} - \rho_\mathrm{s,0})e^{-t/\tau}$, where  $\rho_\mathrm{s,0}$ is the density of cells at the interface at t = 0 min, $\rho_\mathrm{s,\infty}$ is the cell density at long time (t $\to \infty$), and $\tau$ is the characteristic time constant of adhesion. The time constants extracted from the fits are \INKD{4 $\pm$ 1 min, 12 $\pm$ 1, and 16 $\pm$ 3 min for motile, nonmotile (CCCP), and nonmotile (mechanically-sheared) bacteria}, respectively; the time constant of nonmotile \textit{H.\ titanicae} is close to the time constant (9 $\pm$ 4 min) determined in our previous study of adhesion of nonmotile \textit{Marinobacter hydrocarbonoclasticus} on 20 $\mu$m droplets.

\par This result indicates that motility enhances the rate of adhesion at the oil-water interface, but does not affect the order of the kinetics. A variety of colloidal adsorption processes also follow Langmuir kinetics.\cite{myerthall1983kinetics, harris1987vitro, pacha2019kinetic} Likewise, first-order Langmuir-type kinetics have been applied to model the adhesion of \textit{S.\ epidermidis} on silicone,\cite{harris1987vitro} of \textit{Actinomyces viscosus} on silica beads,\cite{myerthall1983kinetics} and of \textit{S.\ epidermidis} on titanium alloy.\cite{pacha2019kinetic}

\subsection{Long-time Adhesion}

\par
\begin{figure}[!ht]
\centering
\includegraphics[width=3.25in]{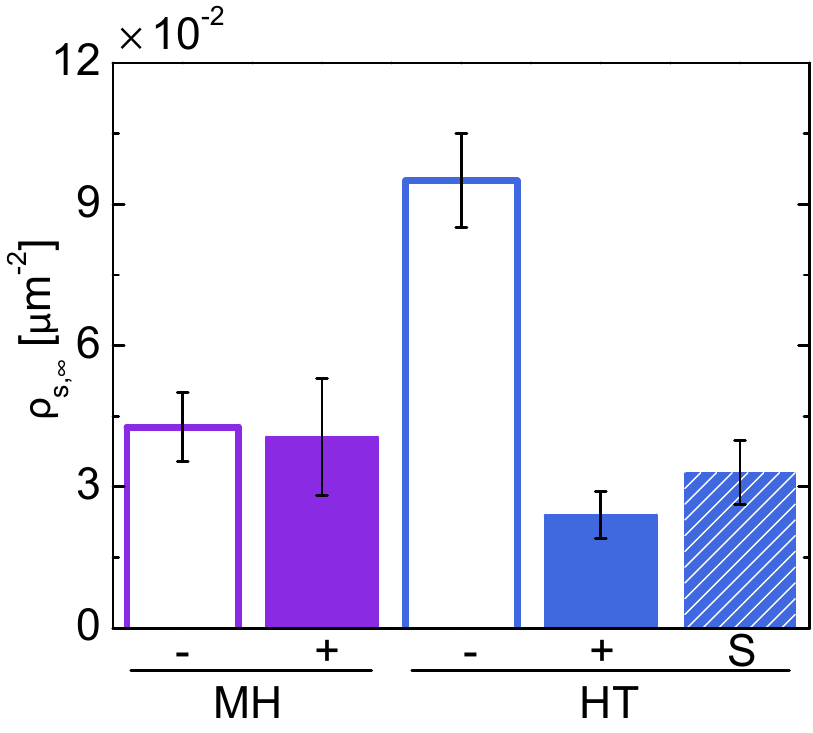}
\caption{\label{fig:CCCPeffect} Effect of motility on adhesion of bacteria to 20 $\mu$m hexadecane droplets. Outer phase is ASW. MH and HT respectively indicate \textit{M.\ hydrocarbonoclasticus} and \textit{H.\ titanicae}. `+' and `-' respectively indicate the presence or absence of 5 $\mu$M of CCCP, which arrests cell motility. \INKD{S} represents data acquired for nonmotile \textit{H.\ titanicae} prepared by mechanically shearing off the flagella. Data acquired for nonmotile \textit{M.\ hydrocarbonoclasticus} suggest that CCCP does not affect the bacterial adhesion through modification of cell or emulsion surfaces. The optical density was 0.2, which corresponds to 5.4$\times10^8$ cells mL$^{-1}$ for \textit{H.\ titanicae} and 3.5 $\times10^8$ cells mL$^{-1}$ for \textit{M.\ hydrocarbonoclasticus}.}
\end{figure}

 In addition to reducing the time needed to reach steady-state adsorption, motility also affects the long-time cell density $\rho_\mathrm{s,\infty} = \rho_\mathrm{s}(t \rightarrow \infty)$. In our experiments, $\rho_\mathrm{s,\infty}$ of motile bacteria is approximately six times greater than that of nonmotile \textit{H.\ titanicae} bacteria (Fig.\ \ref{fig:CCCPeffect}). Our finding is consistent with earlier studies showing that bacteria motility enhances adhesion of bacteria on solid surfaces. \cite{morisaki1999effect, kerchove2008bacterial, tamar2016role, jindai2020adhesion} \INKD{By contrast, the long-time interfacial density of a bacterium that is nonmotile under these experimental conditions, \textit{M.\ hydrocarbonoclasticus}, is not altered by addition of CCCP. Its interfacial density is slightly greater than that of the nonmotile \textit{H.\ titanicae}, consistent with the idea that the chemical structure of the bacterium surface affects interfacial adhesion.}

\par Motility in our experiments is arrested through the addition of CCCP, which may affect the interfacial properties of cells or droplets. To confirm that the difference in $\rho_\mathrm{s,\infty}$ is due to bacterial motility, we characterized the interfacial properties of bacteria and droplets. The zeta potentials (\INKD{Fig.\ \ref{fig:CCCPeffect} and} Table \INKD{1}) of a bacterium that is nonmotile under these experimental conditions, \textit{M.\ hydrocarbonoclasticus}, are \INKD{the same with or without CCCP, within experimental error, although the surface energy is somewhat greater with CCCP}. Furthermore, CCCP does not change the interfacial properties of oil-in-water emulsions, as indicated by the near-constant values of interfacial tension and zeta potential measured for these emulsions (Fig.\ \ref{fig:rho_s}a and Table S1). \INKD{Finally, the surface properties (surface energy and zeta potential) of \textit{H.\ titanicae} remain unchanged in presence and in absence of CCCP.} Together, our results suggest that CCCP alters adhesion through arrest of bacterial motility and not by changing the surface properties of droplets or cells. \INKD{This result suggests that motility is the dominant factor controlling the difference in adhesion.}
 
\par Bacterial adhesion is often considered to be a two-step process. First, a bacterium must diffuse from bulk of liquid suspension to close to the droplet surface. Second, the bacterium must displace the water layer at the interface to access the surface and subsequently adhere on it.\cite{carniello2018physico} The first of these processes is dominated by bacterial transport; the second is additionally affected by the thermodynamics of colloidal adhesion at a liquid-liquid interface.\cite{absolom1983surface, thaveesri1995granulation, dewangan2018adhesion} From thermodynamics, the number of cells adsorbed on a liquid-liquid interface depends on the free energy of adhesion, which depends on the three-phase contact angle and the surface energy of each phase.\cite{power2007time} To determine how motility may affect these processes, we estimate the transport-controlled interfacial cell density for nonmotile and motile bacteria. We assume that bacteria can be modeled as spheres of diameter $1$ $\mu$m. The diffusivity is then given by the Stokes-Einstein equation, $D_\mathrm{nm} = k_B T / 6\pi\eta R_\mathrm{H}$, where, $R_\mathrm{H}$ is the hydrodynamic radius. For a prolate ellipsoid,  $R_\mathrm{H} = \sqrt{a^2-b^2} / \ln{([a+\sqrt{a^2-b^2}]/b)}$, where $a$ and $b$ are the major and minor axis lengths, respectively.\cite{he2003novel} Taking the room temperature as $T =$ 298 K, the viscosity of the medium as that of water, $\eta =$ 0.89 cP, and the major and minor axes of a bacterium as $a = 2$ $\mu$m and $b = 0.6$ $\mu$m, respectively, we estimate $R_H \approx 1$ $\mu$m and $D_\mathrm{nm} \approx$ 0.25 $\mu$ m$^{2}$s$^{-1}$. In our experiments, bacteria are much smaller than the oil droplets (20 -- 35 $\mu$m). Finally, we assume that the adsorption of bacteria is nearly irreversible and that the bulk cell concentration does not deplete over time. With these assumptions and conditions, the density of cells on the droplet surface is given by $\rho_{s}{(t)} = 2C_\mathrm{0}\sqrt{\frac{D_\mathrm{nm}t}{\pi}}$.\cite{van1996kinetics, junji2013diffusion} \INKD{We note that this $t^{1/2}$ scaling is consistent with the short-time scaling of the Langmuir kinetic model. For that model, this scaling can be obtained from statistical rate theory,\cite{rudzinski2008kinetics} which leads to an exponential form of the general kinetic equation.\cite{rudzinski2008kinetics, marczewski2010analysis} On short times, the general kinetic equation scales as $t^{1/2}$.} We choose the characteristic time scale of adsorption to be 3$\tau$, where $\tau$ is the time constant obtained from adsorption kinetics; at this time scale, the surface density is expected to reach 95\% of its long-time limit. For an initial bulk cell concentration of $5.4 \times 10^8$ cells mL$^{-1}$ the interfacial density of nonmotile cells is predicted to be 0.014 cells $\mu$m$^{-2}$. This value is in reasonable agreement with the experimentally-measured interfacial density, which for nonmotile \textit{H.\ titanicae} cells is $0.024 \pm 0.005$ cells $\mu$m$^{-2}$.

\par Motility increases the effective diffusivity of bacteria.\cite{taktikos2013motility, kim2004enhanced, licata2016diffusion} We observe experimentally that peritrichously flagellated \textit{H.\ titanicae} swim using a run-and-tumble mechanism. The diffusivity of a motile bacterium that undergoes runs and tumbles is given by $D_\mathrm{m} = v^{2}\tau_{run} / 3(1-\alpha)$, where $v$ is the mean swimming speed, $\tau_{run}$ is the mean duration of straight runs, and $\alpha$ is the mean value of the cosine of the angle between successive runs.\cite{berg1993random} For successive runs that are uncorrelated in direction, $\alpha$ = 0 and $D_\mathrm{m} = v^{2}\tau_{run}/3$. In our previous study,\cite{dewangan2019rotating} the average swimming speed of \textit{H.\ titanicae} was 10 $\mu$m s$^{-1}$, yielding  $D_\mathrm{m} = 33$ $\mu$ m$^{2}$s$^{-1}$. We again take the characteristic time scale to be $3 \tau = 9$ min. Therefore, the density of motile cells $\rho_{s}{(t)} = 2C_\mathrm{0}\sqrt{\frac{D_\mathrm{m}t}{\pi}}$ is predicted to be 0.081 cells $\mu$m$^{-2}$. \INKD{The gradual decrease in motility over 1 $\--$ 3 min after addition of NaCl may affect the interfacial cell density measurement on shorter time scales. Using $\rho_{s}{(t)} = 2C_\mathrm{0}\sqrt{\frac{Dt}{\pi}}$, \cite{van1996kinetics, junji2013diffusion} we estimate that the interfacial density is at most 10\% and 26\% greater than at t = 6 min and t = 8 min, respectively, compared to that at t = 5 min.} This value is close to the experimentally obtained value of $0.095 \pm 0.010$ $\mu$m$^{-2}$. The reasonable agreement between the predicted and measured densities suggests that motility can increase adhesion by increasing the flux towards the interface. We note, however, that our adhesion curves approximately attain a plateau and do not follow $t^{1/2}$ scaling\cite{van1996kinetics, junji2013diffusion} on long times. This finding suggests that the accessible surface for adhesion becomes saturated after the initial increase; in this context, motility allows the effective packing density to be increased.\cite{kerchove2008bacterial} \INKD{Droplets incubated with \textit{H. titanicae} bacteria at room temperature for 48 hours did not change in size.}

\subsection{Effect of DOSS Concentration on Adhesion of Bacteria to Hexadecane Droplets.}
\par
\begin{figure}[!htbp]
\centering
\includegraphics[width=3.25in]{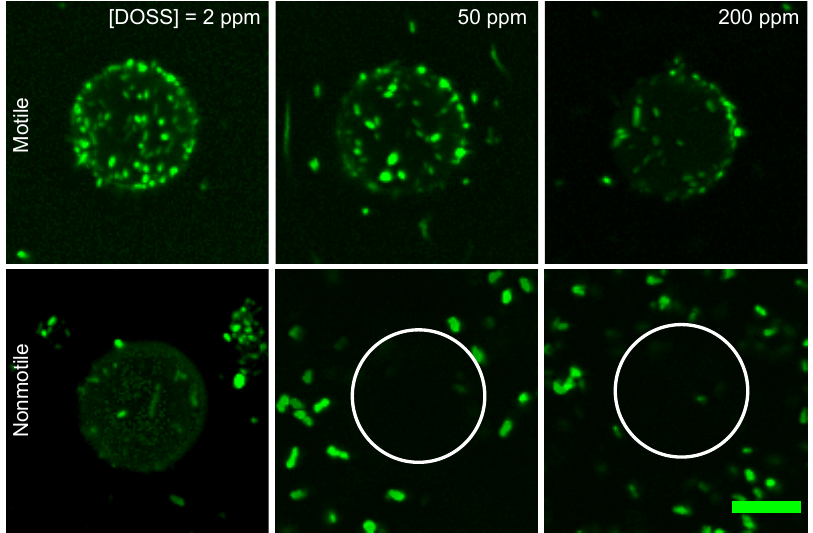}
\caption{\label{fig:imagesDOSS} 2D projections of 3D confocal images of cells adhered on 20 $\mu$m hexadecane droplets for motile bacteria (top panel), and nonmotile bacteria (bottom panel) for [DOSS] = 2 (first column), 50 (second column), and 200 ppm (third column). Scale bar is 10 $\mu$m. White circle represents the location of droplet for clarity.}
\end{figure}

 We applied our imaging method to investigate the effect of surfactant concentration on adhesion of motile and nonmotile \textit{H.\ titanicae} to hexadecane droplets (Fig.\ \ref{fig:imagesDOSS}). We chose as the surfactant DOSS, a major component of the Corexit dispersant used in oil spill response.\cite{place2010role} Confocal micrographs revealed that the number of cells adhered to the oil-water interface decreases as the surfactant concentration is increased (Fig.\ \ref{fig:imagesDOSS}) Earlier studies have reported conflicting trends for bacterial adhesion at oil-water interfaces in the presence of surfactants.\cite{kaczorek2011uptake,  zhang2013influences, lin2017effects} A microbial adhesion to hydrocarbon (MATH) assay showed that \textit{Sphingomonas} \textit{spp.} GY2B adhesion to oil phase decreases with increases in rhamnolipid concentration. \cite{lin2017effects} In contrast, \textit{Klebsiella oxytoca} PYR-1 cell adhesion to oil phase increases concomitant with Tween 20 and Tween 40 concentration up to the critical micelle concentration (CMC). \cite{zhang2013influences}

\par
\begin{figure*}[!ht]
\centering
\includegraphics[width=7in]{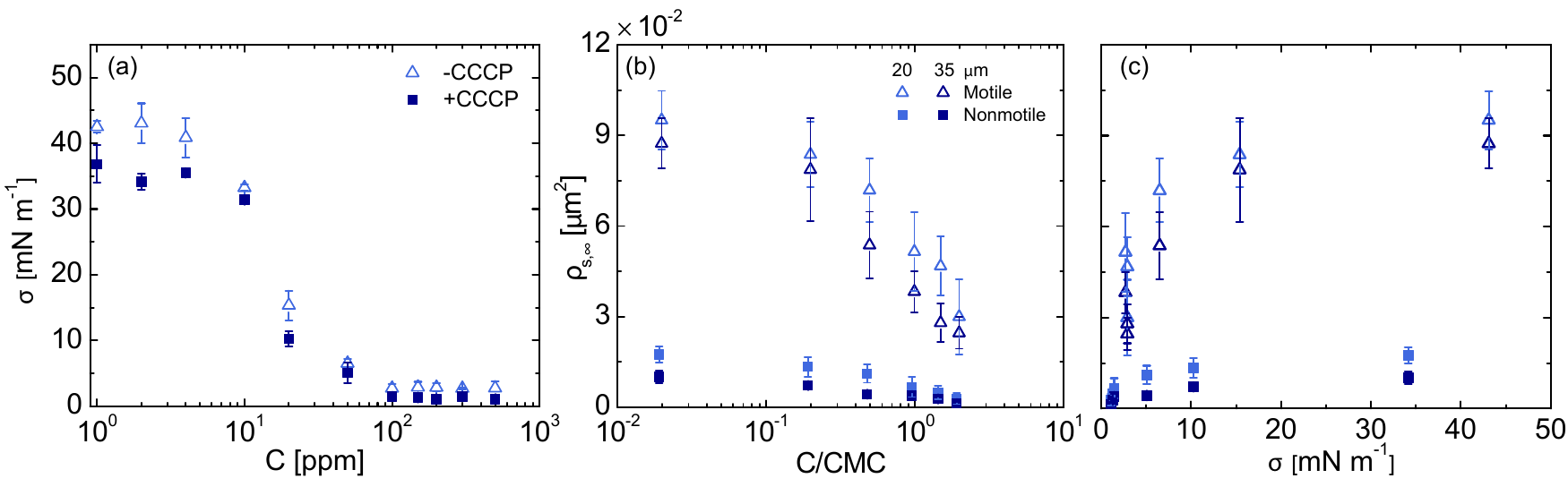}
\caption{\label{fig:rho_s} (a) Interfacial tension of hexadecane/ASW in absence and in presence of 2.5 $\mu$M CCCP as a function of DOSS concentration measured using pendant drop method. (b) and (c) Long-time interfacial density $\rho_\mathrm{s,\infty}$ of cells adhered on hexadecane droplets as a function of (a) normalized surfactant concentration C/cmc, and (b) hexadecane/water interfacial tension. Error bars represent the standard deviation calculated from at least 40 observations (at least 10 droplets from minimum four independent bacterial cultures).}
\end{figure*}

 To understand the effects of surfactant concentration on interfacial adhesion, we first measured the interfacial tension of hexadecane in ASW with and without CCCP in the absence of bacteria. The interfacial tension $\sigma$ is approximately constant at low DOSS concentrations and decreases to near zero at high concentrations of DOSS (Fig.\ \ref{fig:rho_s}a). We determine the critical micelle concentration (CMC) of each DOSS/ASW system (with and without CCCP) from the intercept of linear fits of $\sigma$ as a function of [DOSS] at low and high concentrations (Fig.\ \ref{fig:rho_s}a). Addition of CCCP does not markedly alter $\sigma$: the critical micelle concentration (CMC) is 101 ppm in absence of CCCP and 105 ppm in presence of CCCP. The surface concentrations of DOSS in absence and in presence of CCCP are 3.2 $\times10^{-6}$ and 2.2 $\times10^{-6}$ mol m$^{-2}$, respectively. These are close to the values reported in the literature for the surface concentration of DOSS, which range from 1.6 $\times10^{-6}$ to 2.2 $\times10^{-6}$ mol m$^{-2}$ (depending on the concentration of ammonium chloride) for hexadecane in artificial sea water.\cite{dey2012aggregation}

\begin{table*}[t]
\centering
\small
  \caption{Contact angle, surface energy, and zeta potential \INKD{$\zeta$} of bacteria without CCCP (-) and incubated for with one hour incubation in 5 $\mu$M CCCP at room temperature (+). Standard deviations are calculated from two independent bacteria cultures.}
  \label{tbl:BacteriacSE}
  \begin{tabular}{m{3cm} m{1cm} m{2cm}  m{2cm} m{2cm} m{3.5cm} m{2cm}} 
    \hline
    Bacteria species & CCCP & Water [$^{\circ}$] & DIM [$^{\circ}$] & EG [$^{\circ}$] & Surface energy [mN m$^{-1}$] & $\zeta$ [mV]\\
    \hline
    \textit{H.\ titanicae} & -  & $23 \pm 4$ & $\hspace{0.2cm}57 \pm 5$ & $29 \pm 4$ & $64 \pm 1$ & $-51 \pm 2$ \\
     & + & $25 \pm 2$ & $\hspace{0.2cm}64 \pm 7$ & $27 \pm 4$ & $63 \pm 1$ & $-47 \pm 7$\\
    \textit{M.\ hydrocarbonoclasticus} & - & $86 \pm 3$ & $\hspace{0.2cm}87 \pm 4$ & $98 \pm 5$ & $23 \pm 1$ & $-45 \pm 4$\\
     & + & $56 \pm 7$ & $101 \pm 8$ & $73 \pm 1$ & $42 \pm 1$ &   $-39 \pm 4$\\
    \hline
  \end{tabular}
\end{table*}

\par The density of cells at the oil-water interface decreases as the surfactant concentration is increased (Fig.\ \ref{fig:rho_s}b). The interfacial density of nonmotile cells is nearly zero above the CMC. By contrast, $\rho_\mathrm{s,\infty}$ of motile cells above the CMC is nonzero. Interestingly, $\rho_\mathrm{s,\infty}$ for C/CMC $> 1$ is greater than that of nonmotile cells even at the lowest DOSS concentration (2 ppm). We were able to obtain sufficient statistics for droplets of two diameters: 20 $\mu$m and 35 $\mu$m, both of which fall on the low end of the droplet sizes measured during the Deepwater Horizon oil spill.\cite{li2011monitoring} For motile cells, $\rho_\mathrm{s,\infty}$ for the two droplet diameters is not distinguishable within experimental error (Fig.\ \ref{fig:rho_s}b). For nonmotile cells, $\rho_\mathrm{s,\infty}$ is slightly greater for 20 $\mu$m droplets than for 35 $\mu$m droplets. This result is consistent with our previous work on adhesion of nonmotile \textit{M.\ hydrocarbonoclasticus} to dodecane, in which the highest long-time cell density was obtained on the smallest droplets. \cite{dewangan2018adhesion} The decrease in surface concentration with increasing DOSS concentration is slightly more pronounced for nonmotile bacteria. The percentage decreases in $\rho_\mathrm{s,\infty}$ from low (2 ppm) to high DOSS concentration (200 ppm) are 84 \%, and 68 \% for nonmotile bacteria and motile bacteria, respectively, on 20 $\mu$m, and 88 \% and 72 \% for nonmotile and motile bacteria, respectively, on 35 $\mu$m droplets.

\par The initial adhesion of bacteria is primarily controlled by the interfacial tension and surface energy. \cite{absolom1983surface, zhang2015quantitatively, de2015adhesion, yuan2017surface} Thus, we examined the cell density as a function of the hexadecane-ASW interfacial tension measured in the absence of bacteria $\sigma$ (Fig.\ \ref{fig:rho_s}c). The interfacial densities of both motile and nonmotile cells increase with $\sigma$. This result is consistent with our previous study of the effects of surfactant concentration on adhesion of nonmotile \textit{M.\ hydrocarbonoclasticus} on dodecane droplets.\cite{dewangan2018adhesion} Cell adhesion decreases with increase in surfactant concentration due to an increase in free energy of adhesion.\cite{power2007time, zhang2015quantitatively} The free energy of adhesion increases because $\sigma$ decreases as the surfactant concentration is increased.\cite{absolom1983surface, thaveesri1995granulation} When the interfacial tension is close to zero, $\rho_\mathrm{s,\infty}$ is nearly zero for nonmotile bacteria but nonzero for motile bacteria. These results are consistent with our earlier work on adhesion of nonmotile bacteria to oil droplets, in which we posited that the interfacial cell density was primarily controlled by interfacial tension with additional contributions from electrostatic interactions between surfactant-decorated oil droplets and bacteria.\cite{dewangan2018adhesion}

\section{Conclusions}
\par We investigated the effect of motility on adhesion of bacteria to DOSS-stabilized hexadecane droplets suspended in artificial seawater. The time evolution of the interfacial cell density follows first-order Langmuir kinetics for both motile and nonmotile bacteria. The time constant of adhesion of motile bacteria is smaller than that of nonmotile bacteria, indicating that motility speeds adhesion kinetics. On long time scales the interfacial density $\rho_\mathrm{s,\infty}$ of both motile and nonmotile bacteria approaches a  constant values, and is greater for motile bacteria. This result suggests that motility may enable bacteria to pack more efficiently on the droplet interface. Finally, increasing the concentration of surfactant leads to a decrease in the interfacial tension and a decrease in the $\rho_\mathrm{s,\infty}$ for both motile and nonmotile cells. Although $\rho_\mathrm{s,\infty}$ approaches zero for nonmotile cells at high surfactant concentration, it remains nonzero for motile cells for all concentrations examined. Thus motility may aid bacteria to colonize interfaces with very low interfacial tension.

\par Our results reveal how bacteria motility may enhance adhesion to oil droplets: motile bacteria may (a) adhere at a faster rate and (b) arrange more densely on a surface, as compared to nonmotile bacteria. Because these processes may enhance colonization of dispersed oil, our results suggest that motility may benefit biodegradation during marine oil spills. More broadly, this study contributes to a body of literature\cite{morisaki1999effect, kerchove2008bacterial, jindai2020adhesion} that suggests that bacteria motility may provide an advantage in initial attachment of bacteria to various surfaces.

\section*{Conflicts of interest}
There are no conflicts to declare.

\section*{Acknowledgements}
This research was made possible in part by a grant from The Gulf of Mexico Research Initiative, and in part by the Welch Foundation (E-1869). Data generated for this paper is available on the Gulf of Mexico Research Initiative Information and Data Cooperative (GRIIDC) at https://data.gulfresearchinitiative.org (DOI: 10.7266/4VJXD55E). We thank Kelli Mullane and Dr.\ Douglas Bartlett (UCSD) for providing both bacteria.

\bibliography{ref}

\bibliographystyle{rsc} 

\end{document}